\documentclass[%
 reprint,
 superscriptaddress,
 amsmath,amssymb,
 aps,
]{revtex4-2}

\newtheorem{definition}{Definition}

\usepackage[utf8]{inputenc}
\usepackage{rotating}

\usepackage{amsfonts}
\usepackage{amsmath}
\usepackage{float}
\usepackage{appendix}
\usepackage{graphicx, import}
\usepackage{dcolumn}
\usepackage{bm}
\usepackage{braket}

\usepackage{color}

\makeatletter
\renewcommand*\env@matrix[1][*\c@MaxMatrixCols c]{%
  \hskip -\arraycolsep
  \let\@ifnextchar\new@ifnextchar
  \array{#1}}
\makeatother
\def\env@matrix{\hskip -\arraycolsep
  \let\@ifnextchar\new@ifnextchar
  \array{*\c@MaxMatrixCols c}}

\begin{document}
\newcommand{\updated}[1]{{\color[rgb]{0,0.3,0.6}{#1}}}

\preprint{APS/123-QED}

\title{Not-so-adiabatic quantum computation for the shortest vector problem}

\author{David Joseph}
\affiliation{Electrical and Electronic Engineering Department, Imperial College London.}%
 
 \author{Alexandros Ghionis}
 \affiliation{Physics Department, King's College London}%
 \affiliation{Physics Department, Imperial College London}
 
 \author{Cong Ling}
 \affiliation{Electrical and Electronic Engineering Department, Imperial College London.}%
 
 \author{Florian Mintert}
 \affiliation{Physics Department, Imperial College London}%

\date{\today}

\begin{abstract}
Since quantum computers are known to break the vast majority of currently-used cryptographic protocols,
a variety of new protocols are being developed that are conjectured, but not proven to be safe against quantum attacks. Among the most promising is lattice-based cryptography, where security relies upon problems like the shortest vector problem. We analyse the potential of adiabatic quantum computation for attacks on lattice-based cryptography,
and give numerical evidence that even outside the adiabatic regime such methods can facilitate the solution of the shortest vector and similar problems.
\end{abstract}

\maketitle



\section{\label{sec:intro} Introduction}
The advent of quantum computers heralds an age of new computational possibilities. Two paradigms of quantum computing are gate model and adiabatic quantum computation (AQC): the gate model closely resembles current computing architecture, replacing bits with qubits and retaining control over the smallest building-blocks of the system, and AQC in which the solution for the problem to be solved is encoded into the ground state of a Hamiltonian \cite{Farhi2000QuantumEvolution, Farhi2001AProblem}. Typically one cannot prepare this ground state directly \textemdash otherwise the problem would be straightforward to solve. One therefore begins with a physical system with a Hamiltonian whose ground state one knows how to prepare. The adiabatic theorem then guarantees that a sufficiently slow change from this initial Hamiltonian to the problem Hamiltonian $H_P$ lets the system evolve into the ground state of the latter.

Both paradigms have been demonstrated to be equivalent \cite{Aharonov2004AdiabaticComputation}, though there is not a general way of mapping from one paradigm to the other. The most impactful quantum algorithm discovered thus far is that of Shor for Integer Factorisation and Discrete Logarithm \cite{Shor1997Polynomial-TimeComputer}. Quantum computing is expected to have far reaching consequences, influencing materials science \cite{Babbush2018Low-DepthMaterials}, development of medicines \cite{Aspuru-Guzik2018TheRevolution}, and many other disciplines. Crucially for information security though, large-scale quantum computers \textemdash through application of Shor's algorithm \textemdash will make obsolete most currently operational cryptosystems by solving the underlying mathematical problems that are intractable on classical hardware.

\subsection{Cryptography}
When two parties (Alice and Bob) want to communicate securely over an insecure channel they must use public key cryptography. In this case Alice has a public/private key pair. Anyone can encrypt messages using the public key, but only Alice can decrypt these messages as only she knows the private key. This means Bob can communicate securely without having to already share a secret with Alice. Generally speaking, the public key is derived from the secret key in a manner which is not easily reversible. Public key cryptography is not efficient, and so is mostly used for exchanging an initial secret securely, from which point onwards Alice and Bob can use more efficient private key cryptography, which is not relevant to this paper.

Some of today's most prevalent public key cryptosystems are RSA, Diffie Hellman key exchange, and ElGamal, the security of which rely on the hardness of Integer Factorisation and Discrete Logarithm \cite{Rivest1978ACryptosystems, Diffie1976NewCryptography, Bernstein2009IntroductionCryptography}. These are public key cryptosystems. Reverse engineering the secret key from only the public key and other public information amounts to cracking the cryptosystem, and this is what Shor's algorithm allows us to do for the schemes listed above. 

These developments have necessitated the creation of entire new families of cryptosystems \textemdash and the corresponding field of Post-Quantum Cryptography \cite{Bernstein2009IntroductionCryptography}. The security of each family is based on the hardness of one of a handful of `contender problems'. One of these families is Lattice-Based Cryptography (or LBC). LBC is the most promising area, accounting for almost half of the remaining candidate systems in the NIST Post-Quantum Cryptography Standardization process. Lattice-based constructions derive their security from the Shortest Vector Problem (or SVP, more in Section \ref{sec:Preliminaries}) and other closely related problems \cite{Peikert2016ACryptography}. At present these problems are only conjectured hard, i.e. there is no proof that quantum computers cannot solve them in polynomial time (BQP), there is only an absence of algorithms that can do so either provably or heuristically. It is therefore essential to analyse the security of post-quantum cryptosystems, so as to either verify or disprove their resilience against attacks that may be aided by quantum logical elements.

\subsection{Quantum Computing in  Lattice-Based Cryptography}
So far, most efforts towards quantum attacks on cryptographic protocols have concentrated on gate model quantum algorithms and their applications as subroutines to preexisting algorithms \cite{Laarhoven2013SolvingSearch, Regev2004QuantumProblems} which predominantly fall under either `sieving' or `enumeration'. Sieving takes a large basket of vectors and iteratively combines them to obtain smaller and smaller vectors, whereas enumeration evaluates all vectors in a ball around the origin. Central to these gate model algorithms is the quantum Fourier transform (QFT). The most popular approach has been the use of Grover search to quadratically speed up search of unsorted lists in these algorithms. In 2015, however, it was observed that Grover search could not be applied to enumeration \cite{Laarhoven2013SolvingSearch}, but recently a quantum tree algorithm \cite{Montanaro2015QuantumAlgorithms} was utilised to achieve square root speed up of lattice enumeration with discrete pruning \cite{Aono2018QuantumPruning}. QFT is also a key component of quantum hidden subgroup algorithms which have also been applied to the shortest vector problem on ideal lattices (these are structured lattices embedded in algebraic number fields) \cite{Ducas2017AdvancesLattices, Cramer2017ShortIdeal-SVP}. 

Solving this lattice problem is in essence backwards engineering a private key from the public key and other public information (i.e. reversing the trapdoor process previously mentioned), hence compromising any cryptosystems based on the hardness of this problem (and other related problems). The Learning with Errors cryptosystem \cite{Regev2005OnCryptography}, and many LBC trapdoor functions \cite{Micciancio2012TrapdoorsSmaller}, for example, can be shown to be at least as hard as solving various lattice problems. The concept behind Learning with Errors is the addition of Gaussian noise to a lattice equation, which is otherwise easy to solve via systems of linear equations. An important innovation was the introduction of the Smoothing Parameter \cite{Micciancio2007WorstCaseMeasures}, which describes how much noise can be added before the structure of the lattice is lost and the problem becomes meaningless. This technique of using noise to obfuscate solutions could make LBC a fruitful field in which to apply optimisation algorithms such as those enabled by AQC. At present the lattice community is still a long way from breaking these cryptosystems, as they tend to use lattices in hundreds, or even thousands of dimensions and the best algorithms at the time of writing scale exponentially in the dimension parameter.

The appeal of focusing on gate-model quantum algorithms is the rigorous complexity analyses that can be performed to give theoretical scaling. So far none of these gate-model algorithms threaten LBC. There has not yet been any work done on adiabatic quantum algorithms for LBC. Even though time complexity is generally difficult to estimate for this class of algorithms, they seem particularly suitable for attacks on LBC for two reasons: firstly, because Lattice problems can be formulated as optimisation problems, as we will demonstrate for a quantum setting; secondly, while a major drawback of AQC is the prohibitive time cost of achieving adiabaticity, this may not be a problem here as, up to a threshold, approximate solutions are also admissible. This is significant as it means it is \textit{not} necessary to achieve adiabaticity, thereby potentially avoiding the major time constraints associated with AQC. In this paper, we therefore employ AQC-style algorithms, but with sub-adiabatic time parameters.

In this work we demonstrate a mapping from the Euclidean norm of a vector to the energy of an ultra-cooled bosonic gas in a potential trap. To do so we use a generalised Bose-Hubbard Hamiltonian to describe the energy of the quantum system. We then present an AQC algorithm for solving one of the central lattice problems and analyse its performance on several instances of low dimensional lattices. 

\subsection{Structure}
Section \ref{sec:Preliminaries} introduces lattices and explains the shortest vector problem for which the algorithm is designed. It then covers the necessities regarding Adiabatic Quantum Computing (AQC). Section \ref{sec:Q_SVP} outlines the Hamiltonian we will use, and then we build the mapping from lattice vector-norms to system Hamiltonian ultimately combining this into one SVP algorithm. In Section \ref{sec:Results} we analyse both analytical scaling and simulation results. 

\section{\label{sec:Preliminaries} Preliminaries}
Vectors are denoted by lowercase bold letters, matrices by upper case bold letters, and Hamiltonians by $H$. The length of a vector is defined in terms of a norm. For a vector $\textbf{x} = (x_1,...,x_N)   \in  \mathbb{R}^N$ we write $\| \textbf{x} \|_p = (x_1^p + ... + x_N^p)^\frac{1}{p}$. Any value of $p \geq 1$ can be taken, but common choices are $p=2$ and the infinity norm with $\| \textbf{x} \|_\infty = \lim_{p \rightarrow \infty} \| \textbf{x} \|_p = \max_i \{ \| x_i \| \}$ the infinity norm $l_\infty$. For any choice of $p$ there are two shortest vectors (as lattices are symmetric about the origin), but as these are the same up to sign, we refer to `the', and not `a' shortest vector, the length of which is denoted $\lambda_1(\mathcal{L})$.

When talking about approximation factors, we say $\gamma = \textrm{poly}  (N)$ if $\gamma$ grows asymptotically as $O(N^k)$ for some constant $k$, and $\gamma = \exp (N)$ if $\gamma$ grows asymptotically as $O(k^N)$ for some constant $k$. Similarly, we say an algorithm takes polynomial time if it requires $\textrm{poly}  (N)$ operations to complete, and exponential time if it requires $\exp (N)$ polynomial time operations to complete.

The dot product of two $N$-dimensional vectors is the canonical inner product on Euclidean space given by $\textbf{x} \cdot \textbf{y} = x_1 y_1 + ... + x_N y_N$. 

\subsection{Lattices}
\begin{figure}[H]
    \centering
    \includegraphics[width=\linewidth]{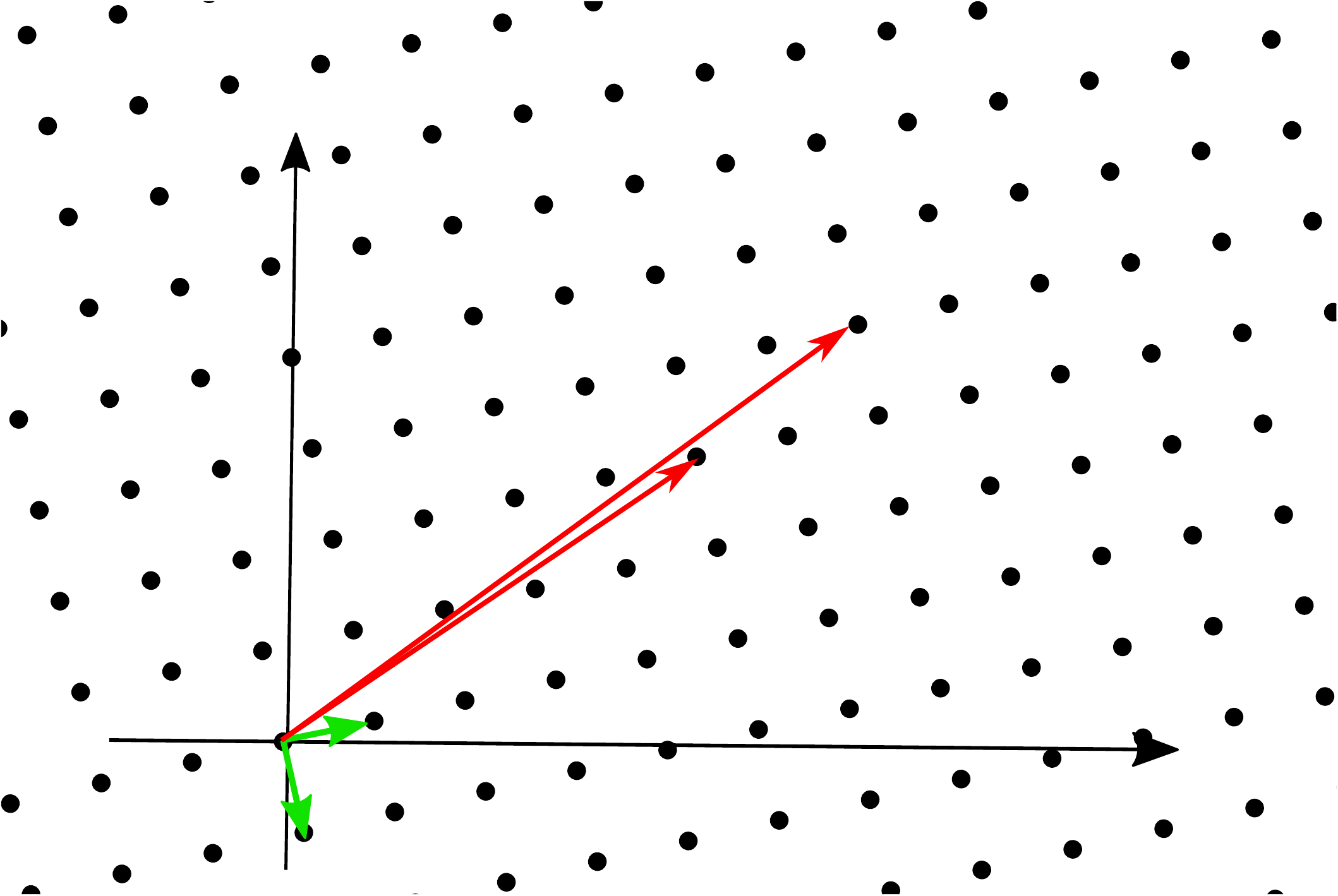}
    \caption{A good basis (green) has short and nearly-orthogonal vectors while a bad basis (red) is the opposite.}
    \label{fig:bases}
\end{figure}

Lattices simply put are a repeating pattern of points in space. In two dimensions, this looks similar to Fig \ref{fig:bases}, which shows that the same lattice can be described by multiple different bases (red arrows and green arrows are just two different bases - there are infinitely many different bases for any given lattice, in fact). The volume of a lattice is equal to the magnitude of the determinant of the basis $\lvert det(\mathcal{L}) \rvert$, and represents the amount of ambient space inside its fundamental parallelepiped \cite{Peikert2016ACryptography}. All bases for the same lattice must therefore have the same determinant, up to sign, but some contain much longer vectors than others, as can be seen by comparing the length of the red arrows with the length of the green arrows in Fig \ref{fig:bases}. A lattice is described by one basis vector for each dimension, and linear combinations of these basis vectors span the entire lattice: in Fig \ref{fig:bases} both red and green bases contain two linearly independent vectors. Mathematically, a lattice is a discrete additive subgroup of $\mathbb{R}^n$. 
\begin{definition}
\label{def:lattice}
    Lattice $L$ is the set of integer combinations of $k$ basis vectors $\textbf{b}_i$, $1 \leq i \leq k$:
    $$L = \Big\{ \sum_{i=1}^k x_i \textbf{b}_i \Big\} = \{ \textbf{x} \cdot \textbf{B} \colon \textbf{x}  \in \mathbb{Z}^k \},$$
\end{definition} 
where the $\textbf{b}_i$ are linearly independent and the lattice is embedded in the ambient space $\mathbb{R}^N$ for some $N \geq k$. The lattice is said to be full rank if $N=k$. Cryptographically, full rank lattices are the most relevant, and also the hardest for a particular dimension ambient space. Because of this, for the rest of this paper we will deal only with full rank integer lattices, i.e. those lattices for which the basis vectors have integer coordinates $\textbf{b}_i  \in  \mathbb{Z}^N$. Throughout the rest of the paper we will treat $\textbf{B}$ as a row basis
\begin{equation}
    \textbf{B} = 
    \left[
    \begin{array}{c}
    \textbf{b}_1 \\
    \vdots \\
    \textbf{b}_N \\
    \end{array}
    \right].
\end{equation}
There is no standardised convention in the cryptographic community (column bases vs row bases) and this choice is generally down to the author's preference. These lattices are in fact (a subset of) Euclidean lattices.

The central problem that we set out to address is the shortest vector problem (SVP) which is simply the task of finding the shortest non-zero lattice vector.
\begin{definition}
\label{def:svp}
    Shortest Vector Problem: Let $\lambda_1(\mathcal{L})$ denote the length of the shortest nonzero vector in a lattice $L$. Given a basis $\textbf{B} = \{ \textbf{b}_1, ..., \textbf{b}_N \}$ describing $L$ find the shortest nonzero vector such that 
    $$\lambda_1(\mathcal{L}) = min\{ \| \textbf{v} \| \colon \textbf{v}   \in  L \backslash \{ \textbf{0} \} \}.$$
\end{definition}
Given a Lattice $L$ determined by a basis $\textbf{B}$, every vector $\textbf{v}$ in the lattice can be described as a linear combination of the basis vectors $\textbf{v} = \textbf{x} \cdot \textbf{B}$ as in Def \ref{def:lattice}. We will call this linear combination $\textbf{x}$ the \textit{coefficient vector} and denote the coefficient vector that achieves the shortest vector $\textbf{x}_{min}$ so that 
\begin{equation}
    \| \textbf{x}_{min} \cdot \textbf{B} \| = \lambda_1(\mathcal{L}).
\end{equation}
In the following we will denote the coordinates of a coefficient vector $\textbf{x}$ as $x_i$ and the coordinates of $\textbf{x}_{min}$ as $x_{min}^i$ to avoid confusion of subscripts.

A variant of SVP is $\gamma$-approximate SVP.
\begin{definition}
\label{def:gamma_svp}
    SVP$_\gamma$: given a basis $\textbf{B} = \{ \textbf{b}_1, ..., \textbf{b}_N \}$ describing a lattice $L$, find $\textbf{v}$ such that $$\| \textbf{v} \| \leq \gamma \cdot \lambda_1(\mathcal{L}),$$
    where $\gamma =  \textrm{poly}  (N)$.
\end{definition}
Cracking this problem is also conjectured hard, and solving it would be considered fatal for LBC. It is this problem that this work targets as the quantum algorithm from Section \ref{sec:Q_SVP} outputs a distribution over short vectors, as discussed in Section \ref{sec:Results}. As such, even if a quantum algorithm for finding \textit{the} shortest vector is not feasible, finding somewhat short vectors may scale significantly better.

The format of the bases we work with to tackle the Shortest Vector Problem have an important bearing on the speed with which we can accomplish the task. With that in mind we will outline the forms of basis that we utilise in this work.

\textit{A note on lattice bases}: in LBC there is much talk of `good' bases and `bad' bases. It is important to distinguish the two and discuss their significance in solving the central problems and compromising lattice-based cryptosystems. A `good' basis is comprised of short vectors which are approximately orthogonal to each other. The conditions of shortness and orthogonality are essentially the same, but they mean that good bases \textit{already} contain short vectors. In a lattice-based cryptosystem one would generate a good basis as a private key (for example, in NTRU \cite{Jeffrey1998NTRUCryptosystem} and GGH \cite{Goldreich1997Public-KeyProblems}, but not in LWE \cite{Regev2005OnCryptography}), and scramble it (making it `worse' and the vectors less orthogonal) to create a bad basis, which would serve as a public key. The instances that are of interest to us are those of bad bases, from which we hope to derive short vectors.

Two types of bases that are relevant to us are Hermite Normal Form bases, which are useful in that they are upper triangular and allow us to perform some useful manipulation later in the paper, and the LLL-reduced bases which are used as a benchmark and a starting point for many lattice algorithms.

\begin{definition}
\label{def:hnf}

Hermite Normal Form (HNF): For any integer lattice row-basis $\textbf{B}$ of rank $N$ there exists a unique upper triangular basis $\textbf{H}$ which satisfies the following conditions:
\begin{itemize}
    \item $\textbf{H}_{ij} = 0$ for $i > j$
    \item $\textbf{H}_{ii} > 0$ for all $i$
    \item $0 \leq \textbf{H}_{ji} \leq \textbf{H}_{ii}$ for $j < i$, or the columns above each pivot (the first nonzero entry from the left) $\textbf{H}_{ii}$ are reduced modulo the pivot
\end{itemize}
\end{definition}
HNF bases form a good starting point for some of the work in Appendix \ref{app:band}. They are generally quite bad bases but have some nice properties which we will use.

LLL-reduced bases \cite{Lenstra1982FactoringCoefficients} are better. The LLL algorithm runs in polynomial time, reducing bad bases to better ones; it outputs vectors which are exponentially larger than those which would be considered solutions to SVP$_\gamma$, but is used as a benchmark in LBC cryptanalysis.

\begin{definition}
\label{def:lll}

LLL-reduced basis: given a basis $\textbf{B} = \{ \textbf{b}_0, ..., \textbf{b}_N \}$, define its Gram-Schmidt orthogonal basis $\textbf{B}^* = \{ \textbf{b}_0^*, ..., \textbf{b}_N^* \}$ and the Gram-Schmidt coefficients 
$$\mu_{i,j} = \frac{\langle \textbf{b}_i, \textbf{b}_j^* \rangle}{\langle \textbf{b}_j^*, \textbf{b}_j^* \rangle}.$$
Then the basis $\textbf{B}$ is LLL-reduced if there exists $\delta   \in  (0.25,1]$ such that:
    \begin{enumerate}
        \item Size Reducetion: For $1 \leq j < i \leq N \colon \lvert \mu_{i,j} \rvert \leq 0.5$
        \item Lovász condition: For $k = 1,...,N \colon \delta \| \textbf{b}^*_{k-1} \|^2 \leq \| \textbf{b}_k^* \|^2 + \mu^2_{k,k-1}  \| \textbf{b}^*_{k-1} \|^2 $.
    \end{enumerate}
\end{definition}
LLL-reduced bases are not unique - there are potentially many different bases satisfying these conditions for any given lattice. In this respect they are different from HNF bases, which are unique for each lattice.
\begin{definition}
\label{def:gram}
    The Gram matrix $\textbf{G}$ of a row basis $\textbf{B} = \{ \textbf{b}_1, ..., \textbf{b}_N \}$ is given by
    $$\textbf{G} = \textbf{B} \cdot \textbf{B}^T.$$
\end{definition}
The Gram matrix will be enough to define $H_P$ entirely, as $\textbf{G}_{ij}$ is the dot product $\textbf{b}_i \cdot \textbf{b}_j$ and so will be used regularly in the following work.

\subsection{Hamiltonian Evolution}
\begin{figure}[H]
    \centering
    \includegraphics[width=\linewidth]{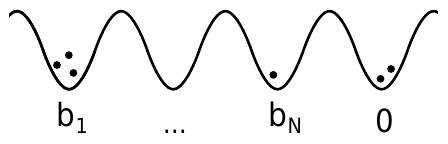}
    \caption{The energy of the system of particles trapped in a potential landscape depicted above is proportional to the square of the Euclidean norm of the vector $\textbf{v} = (3,...,1,2) \cdot \textbf{B}'$.}
    \label{fig:optical lattice}
\end{figure}

The model system that we will use in the following for AQC algorithms is based on the Bose-Hubbard Hamiltonian describing bosonic particles in potential landscapes with sufficiently well pronounced minima that can be identified as sites \cite{Gersch1963QuantumBosons, Bloch2005UltracoldLattices, Jaksch1998ColdLattices, Greiner2002QuantumAtoms.}, and in practice these sites often form a periodic structure, as depicted in Fig \ref{fig:optical lattice}.

The explicit Hamiltonian
\begin{equation}
\label{eq:BH1}
    H_t = f(t) H_0 + g(t) (H_I +   H_S )
\end{equation}
is comprised of a tunnelling term $H_0$, an interaction term $H_I$, and an onsite-energy term $  H_S $, where $S$ represents `site'. Here the sum of the interaction and onsite Hamiltonians give the problem Hamiltonian, $H_P = H_I + H_S$. The annihilation ($a_i$), and creation ($a_i^\dagger$) operators
\begin{equation}
\begin{split}
    &a_i \ket{n_i} = \sqrt{n_i} \ket{n_i-1}\\
    &a_i^{\dagger} \ket{n_i} = \sqrt{n_i + 1} \ket{n_i+1}\\
    &a_i^\dagger a_i = \hat{n}_i,
\end{split}
\end{equation}
decrease, or increase the particle number at site $i$ by one. These operators are used to define the tunnelling term
\begin{equation}
\label{eq:tunnelling_Hamiltonian}
    H_0 = - \sum_{i} ( a_i a_{i+1}^{\dagger} + a_{i+1} a_i^\dagger ),
\end{equation}
where index $i$ runs over the sites in the potential landscape. It will later become clear that the number of sites corresponds directly to the dimension of the lattice in which a short vector is sought.

The interaction Hamiltonian
\begin{equation}
    H_I = \sum_i v_{ii} \hat{n}_i (\hat{n}_i -1) + \sum_{i \neq j} v_{ij} \hat{n}_i \hat{n}_j,
\end{equation}
and onsite Hamiltonian 
\begin{equation}
    H_S  = \sum_{i} \mu_i \hat{n}_i,
\end{equation}
together define the problem Hamiltonian and the interaction constants $v_{ij}$ and onsite energies $\mu_i$ will be determined by the underlying Euclidean lattice. In particular, it will be essential not only to consider onsite interactions $\sum_i v_{ii} \hat{n}_i (\hat{n}_i -1)$ and interactions between neighbouring sites, but also long-range interactions.

With the choice of $f(0)=1$ and $g(0)=0$, the system Hamiltonian contains initially only the tunnelling term. It has comparatively simple eigenstates, and the system can thus be initialised in its ground state. As soon as the values of $f(t), g(t)$ differ from their initial values, the system state will start to evolve in time, but the system will remain in the instantaneous ground state of its current Hamiltonian if the values of $f(t), g(t)$ change sufficiently slowly \cite{Born1928BeweisAdiabatensatzes}. In general, $f, g$ need not be continuous, but to achieve adiabaticity, continuity is necessary (but not sufficient). Given the validity of such adiabatic dynamics, the system will thus end up in the ground state of the problem Hamiltonian $H_I +  H_S $ at the final point in time with $f(T)=0$ and $g(T)=1$.

Since besides the requirement of sufficiently slow changes, there are no further restrictions on $f(t)$ and $g(t)$ there is a continuum of possible sweeps. Knowledge of the spectrum of the underlying Hamiltonian could be used to find functions that make the adiabatic approximation particularly good. Since, however, the AQC should be applicable to the case in which finding this spectrum is beyond computational capabilities, we will not assume any suitably chosen functions, but simply a linear sweep
\begin{equation}
\label{eq:linear evolution}
    H_t = \Big( 1-\frac{t}{T} \Big) \cdot H_0 + \frac{t}{T} \cdot (H_I +   H_S )
\end{equation}
throughout the rest of this paper.

\section{\label{sec:Q_SVP} Quantum Algorithm}
In this section we formulate the quantum SVP algorithm and detail the mapping from vector norms to the Hamiltonian of Eq \eqref{eq:BH1}.

\subsection{Problem Hamiltonian to $l^2$ norm}
The interaction term has an explicit distinction between the interaction $v_{ii} \hat{n}_i (\hat{n}_i - 1)$ of particles at the same site and the interaction $v_{ij} \hat{n}_i \hat{n}_j$ between particles at different sites. This distinction is necessary because $\hat{n}_i$ particles interact only with the remaining $\hat{n}_i - 1$ particles at the same site $i$, whereas $\hat{n}_i$ particles at site $i$ interact with all $\hat{n}_j$ particles at site $j$. On the other hand, there is the onsite interaction term, and the onsite energies can always be chosen such that they compensate for the difference between the onsite and offsite interactions, i.e. such that
\begin{equation}
    H_I +  H_S  = \sum_{i,j} \Tilde{v}_{ij} \hat{n}_i \hat{n}_j
\end{equation}
which maps to the $l^2$ norm of a vector in a natural fashion.

A vector $\textbf{v}   \in  L \backslash \{\textbf{0}\}$ can be written as a unique combination of the basis vectors $\textbf{b}_i$
\begin{equation}
\begin{split}
     & \textbf{v} =  x_1 \textbf{b}_1 + \ldots + x_N \textbf{b}_N
\\
     & \textbf{v}  = \textbf{x} \cdot \textbf{B}.
\end{split}
\end{equation}
Remembering that $\textbf{B}$ is a row basis for $L$, where $\textbf{b}_i$ are the rows. Expanding the square of the Euclidean norm of this vector term by term we have
\begin{equation}
\begin{split}
    \| \textbf{v} \|^2 = &(x_1 \textbf{B}_{11} + \ldots + x_N \textbf{B}_{1N})^2\\ 
    &+ \ldots \\
    &+ (x_1 \textbf{B}_{N1} + \ldots + x_N \textbf{B}_{NN})^2 .
\end{split}
\end{equation}
This can be expressed as
\begin{equation}
\begin{split}
    \| \textbf{v} \|^2 = &(x_1^2 \textbf{b}_1 \cdot \textbf{b}_1 + \ldots + x_N^2 \textbf{b}_N \cdot \textbf{b}_N)\\
    &+ 2(x_1 x_2 \textbf{b}_1 \cdot \textbf{b}_2 + \ldots + x_{N-1} x_N \textbf{b}_{N-1} \cdot \textbf{b}_{N}).
\end{split}
\end{equation}

Referring to \eqref{eq:H_P}, this form neatly fits that of the problem Hamiltonian $H_P$ with the identification of $\Tilde{v}_{ij}$ with the scalar product $\textbf{b}_i \cdot \textbf{b}_j$ of two basis vectors, and of $\hat{n}_i$ with the integer expansions coefficients $x_i$. Generally, an experimentally observable expectation value of particle number at any given site does not need to be an integer, but at the end of the algorithm, where the tunnelling term is vanishing, any local particle number is indeed well defined without quantum fluctuations, so that the identification of $\hat{n}_i$ with $x_i$ is justified.

In terms of Def \ref{def:gram}, the problem Hamiltonian for the AQC thus reads
\begin{equation}
\label{eq:H_P}
    H_P = \sum_{ij}^N \textbf{G}_{ij} \hat{n}_i \hat{n}_j,
\end{equation}
where there are $N$ sites, the energy of a single particle at site $i$ corresponding to the length of basis vector $\textbf{b}_i$. All the interaction constants $\textbf{G}_{ij}$ are defined by the basis $\textbf{B}$ and any non-negative number of particles can be found at any site, subject to availability of particles. Running this algorithm with $K$ particles, they could theoretically occupy the sites in any non-negative combination summing to $K$, resulting in a Hilbert space of $D$ Fock states \cite{Raventos2017ColdDiagonalization}, where $D$ is
\begin{equation}
\label{eq:hilb size}
    D = \frac{(K+N-1)!}{K!(N-1)!},
\end{equation}
and the Fock state with the lowest energy $H_P$ has a configuration of particles that when interpreted as the coefficient vector $\textbf{x}$, gives the shortest possible vector norm $\| \textbf{v} \|$ under the constraint that $x_i \geq 0, \ \forall  \ 1 \leq i \leq N $.

See Appendix \ref{sec:trivial_example} for a worked-though example demonstrating the theory outlined up to this point.

\subsection{Adaptation to Negative Coefficients}
The mapping so far transforms the Euclidean norm squared of general lattice points $\textbf{v} = \textbf{x} \cdot \textbf{B}$ into the problem Hamiltonian energy where the $x_i$ is the number of particles at each site $\hat{n}_i$. The dilemma that this presents is that this only permits non-negative values for each of the $x_i$. This is not a problem, however, as we show next how to modify the physical system such that $H_P$ generalises so as to return solutions that relate to negative $x_i$ values.

The solution that we propose is to add $Nm$ extra particles to the system ($m$ particles for each site), and then by a change of variables use these particles as an offset, thereby permitting negative coefficients $x_i$. The coefficients $x_i$ can now take values as low as $-m$, which occurs if the particle number at site $i$ is zero. If there are $m$ particles at site $i$ then $x_i=0$ and so on. The new particle number can be written $n_i = \hat{n}_i + m$. Denote this new problem Hamiltonian $H_P'$. Upon substituting this change of variables into Eq \eqref{eq:H_P} the new problem Hamiltonian $H_P'$ becomes
\begin{equation}
\begin{split}
    H_P' &= \sum_{ij} \textbf{G}_{ij} ( \hat{n}_i + m)( \hat{n}_j + m)  \\
    H_P' &= H_P + 2m \sum_i \hat{n}_i \Big( \sum_{j} \textbf{G}_{ij} \Big) + m^2 \sum_{ij} \textbf{G}_{ij}.
\end{split}
\end{equation}

To obtain the desired minimisation of the problem Hamiltonian $H_P$ from $H_P'$ the onsite energy needs to be reduced by a function of the column sum of the interaction matrix. The final term, being constant, can be corrected at a later stage so as to return the correct short lattice vectors but would not affect the energy spectrum of the Hamiltonian (other than a constant shift) or, consequently, which configuration of particles minimises the system energy.

To guarantee that the shortest vector lies in the solution set, the offset $m$ must be larger than the infinity norm of the coefficient vector $\textbf{x}_{min}$. That is, $m \geq \| \textbf{x}_{min} \|_\infty$ where $\textbf{x}_{min} \cdot \textbf{B} = \lambda_1(\mathcal{L})$.

\subsection{Multi-Run Quantum SVP}
Above we have defined a mapping from the Euclidean length of a vector to the energy of an ultra-cooled bosonic gas trapped in a potential landscape. But choosing the parameters for total particle number $K$ and length of time evolution $T$ then performing the quantum algorithm are not enough, on their own, to obtain the shortest vector. One run of the algorithm described above contains $K$ particles, but the Fock states in the solution space may not correspond to the required linear combination $\textbf{x}_{min}$.

Take for example a 2D lattice basis for which the shortest vector is determined by $\textbf{x}_{min} = (3,0)$, then $\lambda_1 (\mathcal{L})$ is found by
\begin{equation}
    \| (3,0) \cdot \textbf{B} \| = \lambda_1(\mathcal{L}).
\end{equation}

To obtain the shortest vector using the algorithm detailed above (assuming no prior knowledge about $\textbf{x}_{min}$, and for simplicity setting offset $m=0$) one would first run with particle number $K=1$, then with $K=2$, $K=3$ and then possibly repeat a few more times to be sure the shortest vector has indeed been found. In this way, a search for the shortest vector consists of running the algorithm many times, each time incrementing $K$ by 1 until confident that there are no shorter vectors to be found. The output of this algorithm, if performed adiabatically, will be a collection of coefficient vectors $\textbf{x}$ and the resulting lattice vectors $\textbf{v}=\textbf{x}\cdot \textbf{B}$, each of which return the shortest lattice vector possible for a particular choice of $K$, and among these samples will be the sought after shortest vector of length $\lambda_1(\mathcal{L})$.

The Multi-Run algorithm ensures that with well chosen $m$, $K_{max}$, and sweep times the $\lambda_1(\mathcal{L})$ will definitely be correctly identified. This is after performing the sweeps for particle numbers $K_i = Nm + i$ up to $K_c=K_{max}=Nm+c$. The number of runs required is therefore $c$, which should approximate the absolute coefficient vector sum $\lvert \sum_{i=1}^N x_i \rvert$. As can be seen in black in Fig \ref{fig:kgrowth}, this grows linearly in $N$, meaning $O(N)$ repetitions would be expected in order execute Multi-Run, with each repetition having a different number of particles and being carried out adiabatically.

In the next part we present a more eloquent all-in-one algorithm where many different runs from this algorithm are combined into one larger run. It offers an $O(N)$ improvement in space but at the cost of $\lambda_1(\mathcal{L})$ no longer corresponding to the ground state, but instead to the first excited state.

Appendix \ref{sec:trivial_example} illustrates what one of the runs would look like for a 2D lattice with no offset ($m=0$). 

\subsection{Single-Run Quantum SVP}
The aim of this algorithm is to generalise Multi-Run into one overarching algorithm (Single-Run) that encompasses all of the repetitions executed during the Multi-Run algorithm. Whereas before the coefficient vectors (in 2D) $(1,0)$ and $(1,1)$ could be obtained only from separate runs, now the aim is to include all possible coefficient vectors in one solution space. Instead of repeating sweeps with a different particle number $K$ many times, only one sweep is performed, the solution space of which includes all possible solutions from the Multi-Run version.

What we propose is to introduce an extra site to the potential landscape, corresponding to the zero vector. Label this site $N+1$. This new site should act as a `particle reservoir' and NOT influence the energy of the system directly. For the Euclidean lattice, one appends the zero vector to the basis $\textbf{B}$, as defined in Eq\eqref{eq:new basis}. Denote the particle number for the Single-Run version $K_S$. If the process is run with $K_S \geq K_{max}$ total particles, then the set of configurations where no particles are in site $N+1$ correspond to one run of the previous algorithm with $K_S$ particles; the set of configurations with one particle in site $N+1$ correspond to a run of the previous algorithm but with $K_S-1$ particles, and so on. It is important to remember that the ground state is no longer the shortest vector, but the zero vector - $m$ particles in the first $N$ sites and all the remaining particles in site $N+1$ returns $\textbf{0}$ which has the lowest energy \textemdash and so an adiabatic evolution is no longer desirable. We analyse the implications of this in the following sections.

Accordingly, the Single-Run problem Hamiltonian $H_P'$ looks as follows:
\begin{equation}
\label{eq:ult_Hp}
    \boxed{H_P' = \sum_{ij}^{N+1} \textbf{G}_{ij}' ( \hat{n}_i + m) ( \hat{n}_j + m).}
\end{equation}
where $\textbf{G}_{ij}'$ is defined as $\textbf{B}' \textbf{B}'^T$, for
\begin{equation}
    \label{eq:new basis}
\textbf{B}' =
\left[
\begin{array}{c}
\textbf{b}_1 \\
\vdots \\
\textbf{b}_N \\
\hline
\textbf{0}
\end{array}
\right].
\end{equation}
The problem Hamiltonian in Eq \eqref{eq:ult_Hp} will be the one used for the rest of the paper, unless otherwise stated, including all numerical simulations. 

The size of the Hilbert space $D_S$ for Single-Run Quantum SVP (letting $K_S=K_{max}$ to make the two modes of computation directly comparable) is, either by use of the hockey-stick identity or by direct application of Eq \eqref{eq:hilb size}
\begin{equation}
\label{eq:hilb size single run}
    D_S = \frac{(K_S+N)!}{K_S!N!} = \sum_{K=0}^{K_{max}} \frac{(K+N-1)!}{K!(N-1)!}.
\end{equation}

Having added an extra `particle reservoir' site and accordingly appended $\textbf{0}$ to the basis, the lowest energy state of the problem Hamiltonian is the unwanted \textbf{0} state, but the energy of the first excited state will correspond to $\lambda_1(\mathcal{L})$.

\section{\label{sec:Results} Results}
Little is known analytically about the time scaling for adiabatic quantum algorithms, beyond a worst case energy gap dependence of $1/\Delta^3$ \cite{Jansen2007BoundsComputation} \textemdash with $\Delta$ denoting the minimum energy gap between ground state and second lowest eigenstate \textemdash whereas with quantum gate algorithms neat closed form scalings are known for a handful of algorithms, for example Shor's exponential speedup for Integer Factorisation and Discrete Logarithm \cite{Shor1997Polynomial-TimeComputer} and Grover's quadratic speed up for searching unsorted lists \cite{Grover1996ASearch}. For adiabatic quantum optimisation, it is not yet even know if these algorithms run faster than classical optimisation \cite{Lucas2014IsingProblems}. Due to time dependence on the minimum energy gap between $E_0$ and $E_1$ it is usually found that 
\begin{equation}
    T = O(\exp(\alpha N^\beta)),
\end{equation}
though this leaves open the possibility of drastically reducing run time subject to achieving lower values of $\alpha, \beta$ than their classical analogues.

While it is difficult to estimate the time scaling for this algorithm, or even for which parameter regimes this scaling would be optimal (near adiabatic versus much faster sweeps, for example), we can calculate the qubit space requirements (though we do not directly use qubit-based architecture). 

\subsection{Qubit Requirements}
Let us estimate the required system size. This is a function of number of sites and number of particles $K_S$. The former is predetermined (it is $N+1$) but one can choose the latter. The aim is to make the system just large enough (pick $K_S$) so that $\textbf{x}_{min}$ is one of the possible configurations of particles in $N+1$ sites with high probability.

Firstly $m$, the offset, must be chosen. We derive this by taking an estimate for the infinity norm of $\textbf{x}_{min}$. This can be seen in Fig \ref{fig:kgrowth} for HNF bases (to err of the side of safety, as this will give larger values because HNF bases have long vectors). One can see that the average of the infinity norms grows linearly for HNF bases (see the red best fit line), so we approximate $m$ to be some linear function of $N$. This means that $K_S$ is already up to $m(N+1)$ particles, and so $K_S$ scales as $O(N^2)$.

The only other consideration is the absolute value of the sum of the coefficient terms $k = \lvert \sum_{i=1}^Nx_{min}^i \lvert$. Heuristically this grows linearly as shown by the black best fit line in Fig \ref{fig:kgrowth}, and is less than $m$ so can be ignored from this point on, as in the instance that all $m(N+1)$ particles reside in the first $N$ lattice sites (and none in the particle reservoir corresponding to the zero vector) the net coefficient sum would be $m > k$. This is because $Nm$ particles are acting as offset particles, leaving the remaining $m$ as the coefficient sum.

By considering the size of the solution space, we have analytically deduced that the qubit requirements scale as $O(N \log N )$ as shown in Appendix \ref{app:qubit scaling}, which space-wise appears acceptable.

\subsection{Empirical Results}
\begin{figure}[H]
    \centering
    \includegraphics[width=\linewidth]{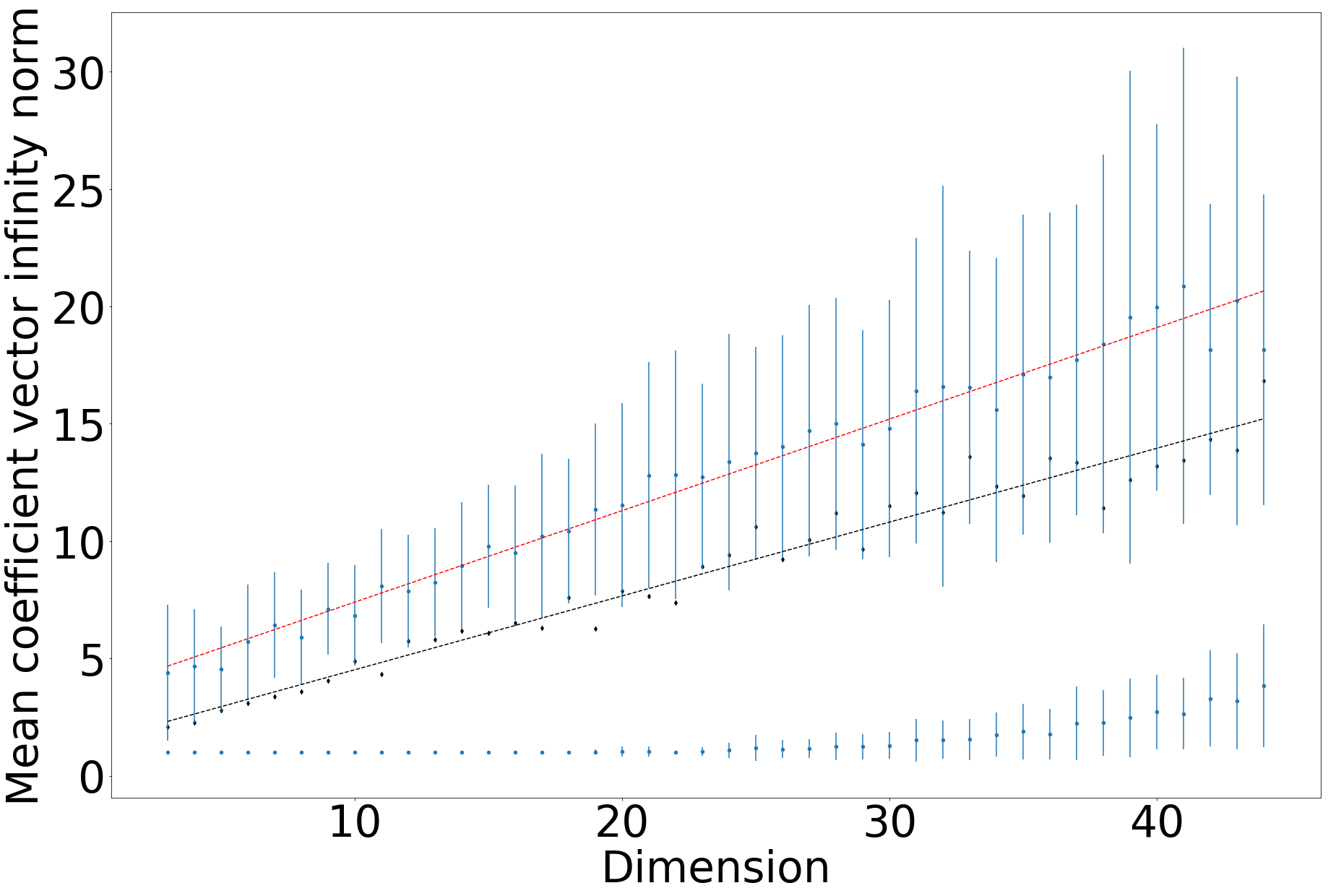}
    \caption{Mean $\| \textbf{x}_{min} \|_\infty$ with error bars averaged over 80 random integer lattices in each dimension $3 \leq N  \leq 45$ calculated on the Hermite Normal Form basis (upper distribution with red dashed best-fit) and LLL reduced bases (lowest distribution). Mean absolute coefficient vector sum ($k= \lvert \sum_{i=1}^N x_i \rvert $) \textemdash with best fit in black dashed line \textemdash demonstrates that $N(m+1)$ particles scales are enough to ensure good chance of containing $\lambda_1$ in the solution set, and also represents expected number of runs in Multi-Run}.
    \label{fig:kgrowth}
\end{figure}

The required particle number $K$ is determined by $ \lvert \sum_{i=1}^N x_{min}^i \rvert $ which must be estimated. This reflects the fact that on average some coordinates of $\textbf{x}$ will be positive, and some negative, cancelling out, but they will rarely cancel out entirely. The growth in the mean of $ \lvert \sum_{i=1}^N x_{min}^i \rvert $ is linear as demonstrated numerically in Fig \ref{fig:kgrowth}, and grows below the estimated offset number $m$. The significance of this is that $m(N+1)$ particles is a generous estimate for $K_S$. This means that taking $K_S = m(N+1)$ gives a good chance of finding $\textbf{x}_{min}$ in the solution set.

To understand this, consider the system with $mN$ particles in the first $N$ sites and $m$ particles in site $N+1$. The potential solutions are the same as those one would get from performing the Multi-Run version with $mN$ total particles. Now with one more particle in the first $N$ sites and one fewer in site $N+1$ the solutions are the same as those from Multi-Run with $mN + 1$ total particles. To ensure there are enough particles in the system, one must be confident of achieving up to $mN +  \lvert \sum_{i=1}^N x_{min}^i \rvert $ particles in the first $N$ sites. That Fig \ref{fig:kgrowth} demonstrates that $ \lvert \sum_{i=1}^N x_i \rvert  < m$ in general means that the total of $m(N+1)$ particles is sufficient to find $\lambda_1(\mathcal{L})$.

The growth in $\| \textbf{x}_{min} \|_\infty$ with respect to Hermite Normal Form bases appears heuristically linear, as can be seen by the red dashed best-fit line in Fig \ref{fig:kgrowth}. 

Another group of cryptographically relevant bases are LLL-reduced bases. These are much better than HNF bases and are easy to obtain. They are often used as a first step in classical SVP routines \cite{Gama2010LatticePruning, Regev2004QuantumProblems}. LLL-reduced bases in low dimensions ($<30$ or so) were so efficient at finding the shortest vector that there were not enough data points to draw sound conclusions from, as can be seen in the lower blue scatter plot in Fig \ref{fig:kgrowth}. We can, however, assert that the LLL-reduced case is upper bounded by the Hermite Normal Form case and so can approximate $m$ to be a term linear in $N$. This is a sound assertion because any lattice basis can be transformed into an HNF basis in polynomial time.

\subsection{Numerical Analysis}
Ideally simulating this quantum SVP algorithm on lattices in many dimensions would give an empirical idea of scaling. Regrettably, simulating quantum systems is computationally very intensive due to the factorial growth of the Hilbert space and so these simulations were only possible for low dimensional lattices. Using the QuSpin python library we were able to simulate problems with Hilbert space sizes of up to ten thousand eigenstates (20 particles in 5 sites). 

Nevertheless it is insightful to consider the distribution over eigenstates (grouped where degenerate) for runs of different time length. We simulated the quantum SVP algorithm on 200, 150 and 100 lattices in two, three, and four dimensions respectively. Using standard `bad' bases from literature does not work well for small dimensions \textemdash both HNF and LLL reduction tend to return maximally reduced bases \textemdash so we generated our own as follows. For each lattice we generated a basis (call this the `good' basis) and then scrambled it by some randomly generated unimodular matrix to obtain a worse basis. The average increase in basis vector length is a factor of 12.06, 10.08 and and 10.07 in dimensions two, three and four respectively under the unimodular transformations. Basis vectors could not be increased by too much otherwise the problems would have become intractable on our hardware. Note also that after generating the Hamiltonians in the QuSpin package we scaled the Hamiltonians in order that they all occupied roughly the same spectrum of eigenvalues. The reason for this is that expanding the energy spectrum significantly increases success probabilities for quantum adiabatic algorithms due to the dependence on minimum energy gap. Scaling Hamiltonians has a similar effect to altering the sweep times which we want to analyse, and it is reasonable to expect that implementers of this algorithm would have access to the same energy spectrum regardless of the problem size. As such, we in a sense `fixed' the spectrum and varied the sweep times to isolate the effect of the parameter $T$.

\begin{figure*}
\includegraphics[width=\linewidth]{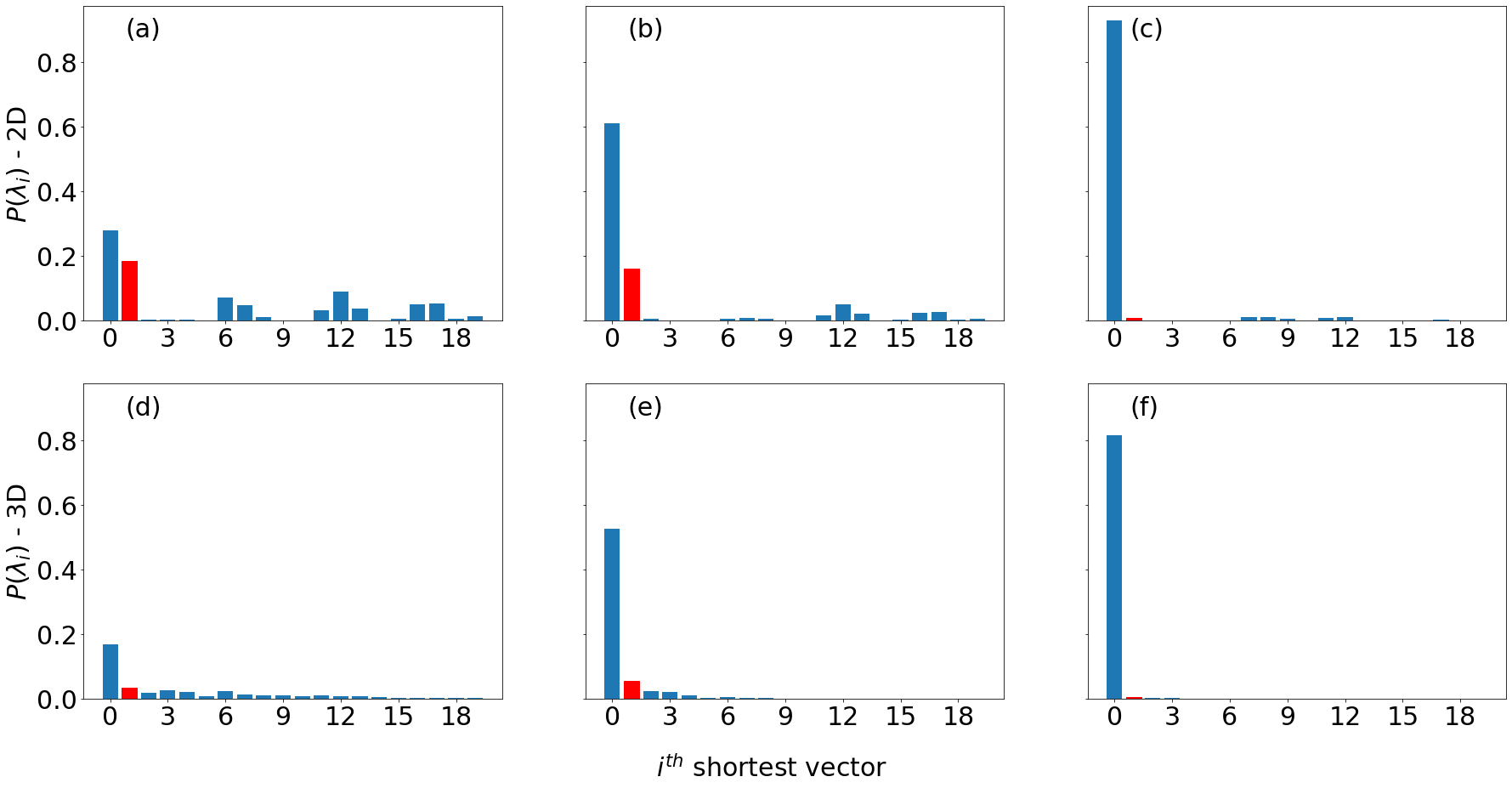}
\caption{\label{fig:23D} Probability of returning $i^{\text{th}}$ shortest vector averaged over 200 instances of 2-dimensional lattices for $T=1, 10, 100$ in plots (a), (b) and (c) respectively; and over 50 instances of 3-dimensional lattices for $T=1, 10, 100$ in plots (d), (e) and (f) respectively.}
    
\end{figure*}

\textit{Mean Distribution over Eigenstates}: 
Fig \ref{fig:23D} shows the averaged results for the Single-Run quantum SVP algorithm of Section \ref{sec:Q_SVP}. Each subplot represents a different choice of parameters, and shows the mean probability of observing the system in an eigenstate corresponding to the zero vector (index $0$), the shortest vector $\lambda_1(\mathcal{L})$ (index $1$, in red), the second, third etc shortest vectors and so on up to the twentieth shortest vector. What is clear to see is a high likelihood of the system being found in low-energy states. For slower sweeps (higher $T$ values) this distribution becomes more concentrated around the lowest-energy states.

Paying particular attention to the red bars (representing the preferred solutions corresponding to $\lambda_1(\mathcal{L})$) one can see that maximising the height of the red bar requires some nuance: sweep too slow and there is too high a chance of attaining the zero vector at an unacceptable time cost; too fast and the system will become excited to much higher energy levels with unacceptably low probabilities of observing the system in very low energy states.

The top row of Fig \ref{fig:23D} displays the final results for parameter choices $N=2$, $m=3$, $T=1,10,100$. This maps to a system of nine particles in three sites. Applying Eq \eqref{eq:hilb size}, the total Hilbert space has 55 eigenstates. The bottom row of Figure \ref{fig:23D} shows the results for parameter choices $N=3$, $m=4$, $T=1,10,100$. This is for a system of sixteen particles in four sites. The Hilbert space now has 969 eigenstates. In both $N=2,3$ slower sweeps result in higher probabilities of the system terminating in the lowest eigenstates, and for $T=100$ there is a very low likelihood of finding the system in anything but the ground state. It should also be observed that as the system size has increased from $N=2$ to $N=3$, and keeping $T$ constant, the probability of recording the system in any given low-energy state decreases.

\subsubsection{Time Sweep Optimisation}
While reliable time bounds for finding ground states of quantum adiabatic algorithms with good enough probability are much sought-after, and the general rule of thumb is `longer is better', thought is shifting. There are instances, for example in the MAX 2-SAT problem, in which slow sweeps perform much worse than fast ones \cite{Crosson2014DifferentAlgorithm}. Here we instead consider faster sweeps and how they relate to algorithms where the exact ground state is not necessarily required.

The problem of observing the ground state of a system after a quantum annealing algorithm has been the subject of much research. In the pursuit of `somewhat' low energy states much less is known. The algorithm outlined in Section \ref{sec:Q_SVP} targets the first excited state, and furthermore, the ground state (corresponding to the zero vector) is of even less use than eigenstates of energy just above $\lambda_1(\mathcal{L})$, as at least these return a short non-zero contender. Bearing this in mind along with the observations from Fig \ref{fig:23D} we thought to examine how targeting low-but-not-ground states versus ground state might differ.
\begin{figure*}
    \centering
    \includegraphics[width=\linewidth]{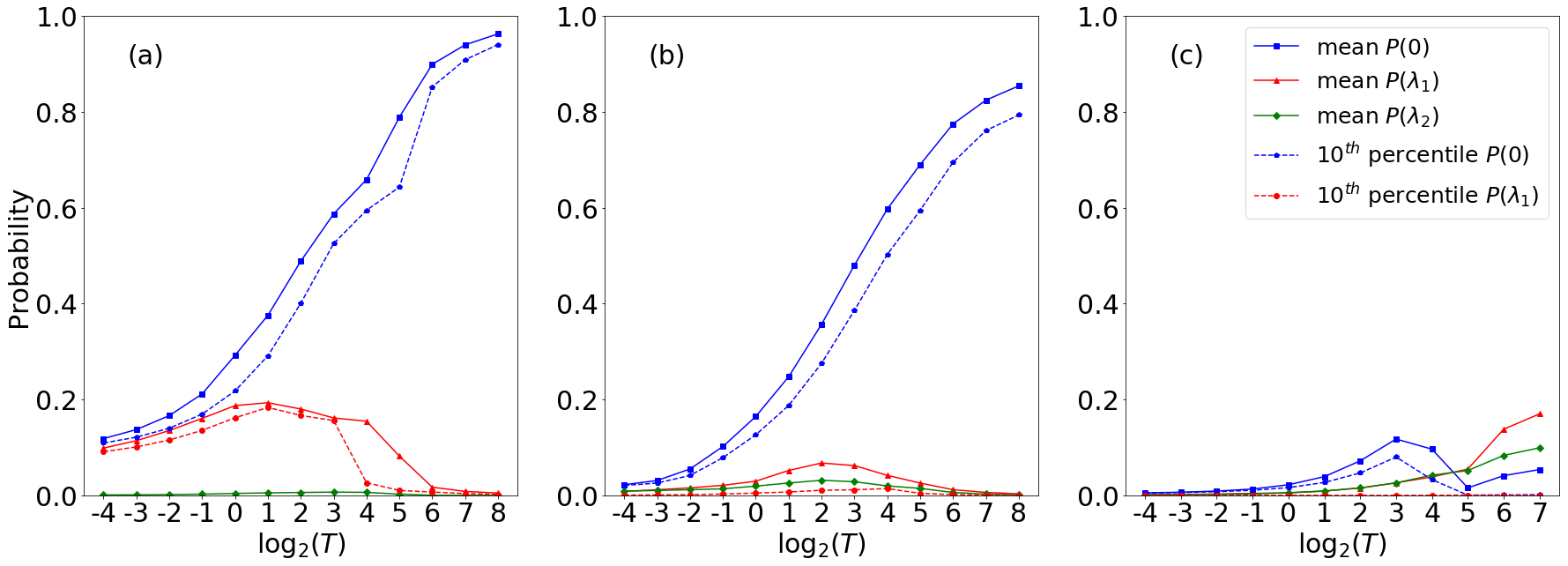}
    \caption{For quantum SVP simulations on (a): 200 2D `bad' lattice bases, (b): 150 3D `bad' lattice bases, and (c): 150 4D `bad' lattice bases, the mean (Solid) and $10^{th}$ percentile (dashed) probabilities of returning $i^{th}$ shortest vector are depicted. Blue is for ground state (zero vector), red corresponds to shortest vector ($\lambda_1(\mathcal{L})$), and green is for the second shortest vector ($\lambda_2(\mathcal{L})$).}
    \label{fig:payoff}
\end{figure*}

\begin{figure*}
\includegraphics[width=0.9\linewidth]{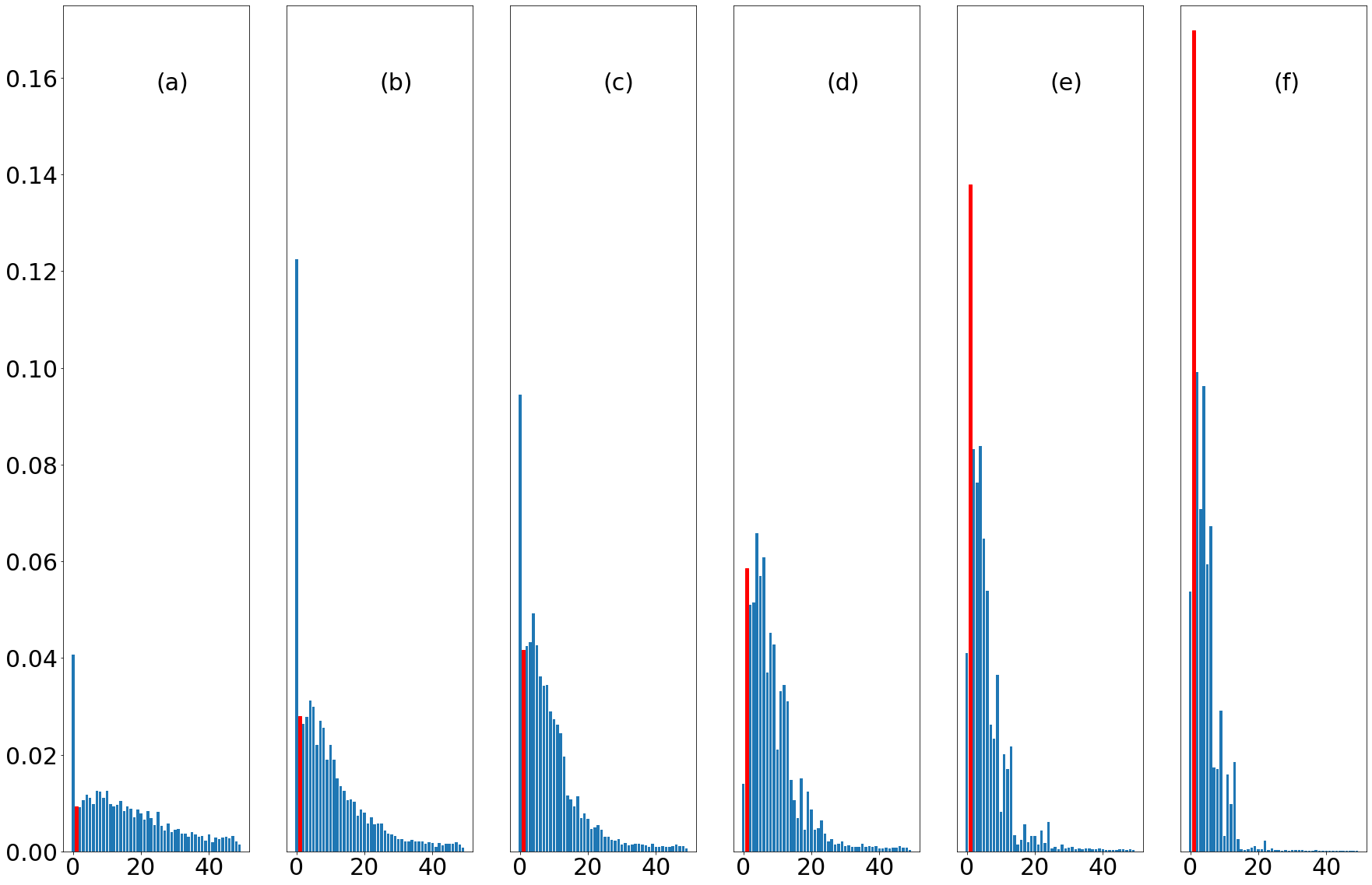}
\caption{\label{fig:4D dist} Mean probability of returning $i^{\text{th}}$ shortest vector across 150 instances of 4-dimensional lattices for sweep lengths $T=4, 8, 16, 32, 64, 128$ in subplots (a), (b), (c), (d), (e), and (f), respectively.}
    
\end{figure*}

The solid lines in Fig \ref{fig:payoff} represent mean probability of returning $\textbf{0}, \lambda_1(\mathcal{L}), \lambda_2(\mathcal{L})$ in blue, red and green respectively. The dashed lines reflect 90\% confidence. 

Looking first at the probability curves there are a few interesting observations to be made. In two and three dimensions, exponentially slower sweeps result in higher probabilities of achieving a final ground state, as might be expected. These blue curves will continue to level off past the right of the axes due to maximum probability of any state being one. But remarkably targetting the first excited state appears to experience almost no success penalty for performing faster sweeps. In fact, the non-monotonicity of the red dashed line indicates that there is some `Goldilocks' zone where evolutions are slow enough to achieve a good distribution over low energy states, but not slow enough that $P(E_0)$ dominates the distribution. In Fig \ref{fig:payoff} this zone appears around $T=2$ for 2D lattices and $T=4$ for 3D lattices.

The case for 4D lattices looks quite different. The solid blue line is overtaken by probabilities for $\lambda_1(\mathcal{L})$ and even $\lambda_2(\mathcal{L})$. To understand better what is happening let us look at the probability distributions at a few different points from plot (c) in Fig \ref{fig:payoff}, which illustrates results on 4D lattices. To this end, Fig \ref{fig:4D dist} presents the same information as that in Fig \ref{fig:23D}, but for more samples and should be looked at closely in conjunction with plot (c) of Fig \ref{fig:payoff}. Again, probabilities corresponding to $\lambda_1(\mathcal{L})$ are highlighted in red. While it is apparent there is some locally different behaviour corresponding to the ground state, Fig \ref{fig:4D dist} shows that as sweeps become slower, probability density continues to accumulate around the lowest energy states, and if slow enough (though beyond our computational capabilities), would concentrate entirely on the lowest eigenstate. This behaviour is particularly promising for the quantum SVP algorithm in higher dimensions as again one can see (subplot (f) provides the best example) a significant concentration of probability around the ten lowest eigenstates (out of $>10,000$) without this distribution necessarily being dominated by the zero vector. Furthermore, the probability of achieving $\lambda_1(\mathcal{L})$ is considerably high relative to surrounding eigenstates for slower sweeps, as demonstrated by the series of red bars in Fig \ref{fig:4D dist} 

In order to explain some of the favourable characteristics it serves to look at a specific instances for the evolution of the system. 

\begin{figure}
    \centering
    \includegraphics[width=\linewidth]{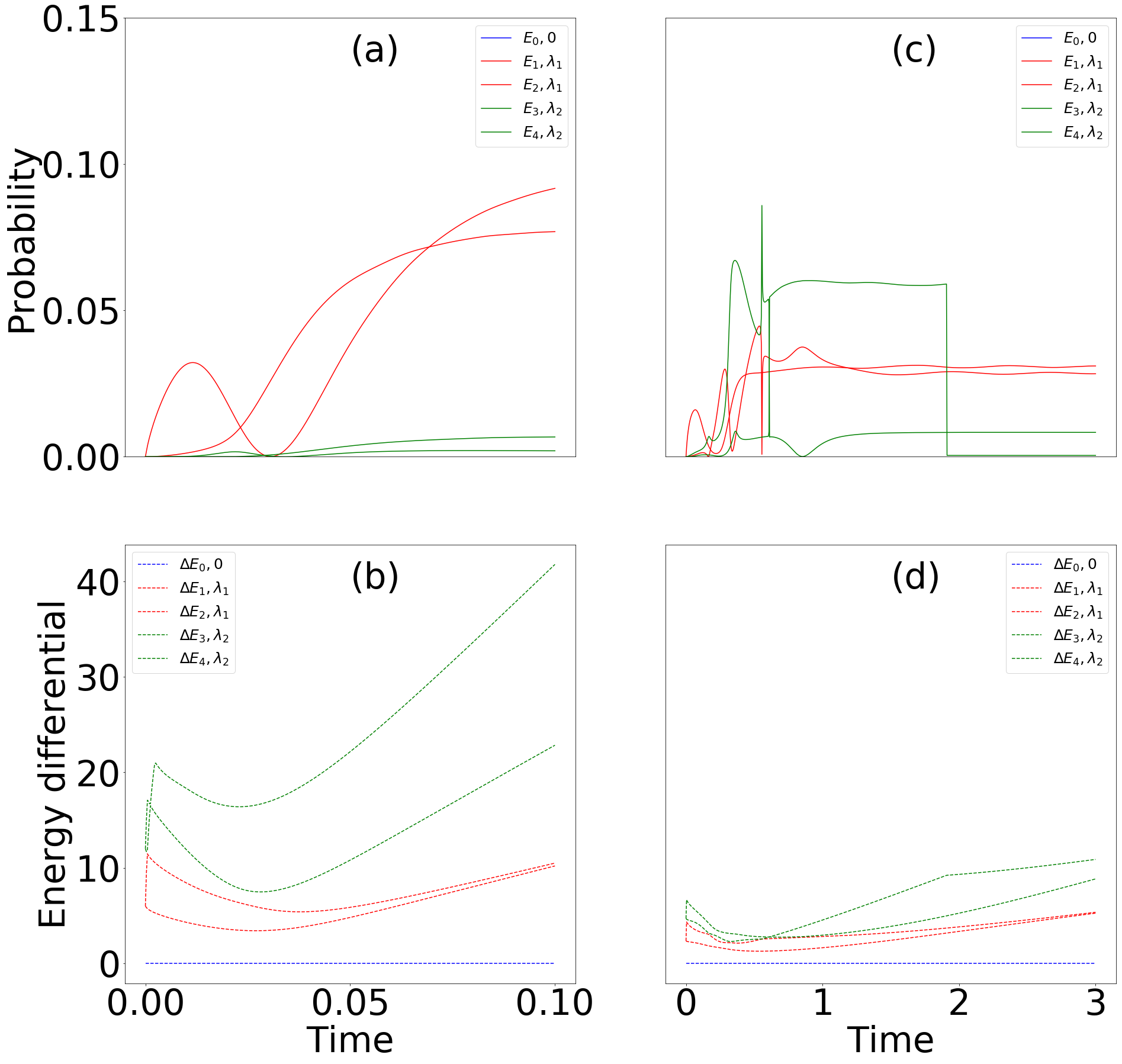}
    \caption{Algorithm sweep on a 2D (left; 0.1 of $T=0.25$) and 3D (right; 3 of $T=4$) lattice basis showing probability of measuring the system in an eigenstate (top) and the corresponding energy differential  (bottom) \textemdash both for only the lowest five energy levels \textemdash defined as $\Delta E_i^t = E_i^t - E_0^t$. Degeneracies should not occur during the intermediate Hamiltonian, though $H_P$ will contain degeneracies, for example those states corresponding to the two shortest vectors.}
    \label{fig:e_levels}
\end{figure}

Fig \ref{fig:e_levels} shows the energy levels and corresponding measurement probabilities for a typical 2D lattice example, run with parameters $T=0.25$, $m=3$.

While the ground state solution relates to the unique zero vector, all other vectors have two associated eigenstates reflecting the symmetry of the lattice about zero (for every $\textbf{v}  \in \mathcal{L}$, its negative $-\textbf{v}$ also belongs to $\mathcal{L}$). This is important as it means that while one excitation is ideal, two excitations still yield $\textbf{x}_{min}$ and hence the shortest vector.

\textit{A note on the trivial solution}: Throughout this piece of work we have assumed the framework of adiabatic quantum computation. The presence of the (useless) zero vector in the solution set means we no longer seek the lowest energy eigenstate. This does not fit the traditional AQC framework, but is advantageous in that it permits faster sweeps. Scaling to larger systems, this could help to circumvent the prohibitive time cost of AQC algorithms. In sub-adiabatic regimes it is foreseeable that shorter sweep times could be employed at the cost of larger $\gamma$ approximations for SVP$_\gamma$ and vice versa.

\section{\label{sec:discussion} Discussion}
We have introduced a quantum optimisation framework to the area of computationally hard lattice problems that may underpin tomorrow's cryptosystems. By examining some interesting properties of an AQC-style algorithm when targeting low-but-not-lowest energy states we have identified the existence of a `Goldilocks' zone for time sweep optimisation. This is particularly exciting for cryptanalysis of lattice-based cryptosystems as the underlying problems often come in approximate \textemdash and not exact \textemdash form, as with SVP$_\gamma$ analysed in this work. Among cryptographers it is thought that the approximate nature of lattice problems strengthens their post-quantum credentials, as the lack of determinism means quantum hidden subgroup algorithms cannot be applied. This `proximity' property, however, may allow sub-adiabatic algorithms for such problems to overcome the costly time requirements of AQC, while still outputting acceptable solutions. Outside of cryptography, it should be observed that for many real-world problems an approximate solution is fine where exact solutions are intractable, and the `Goldilocks' zone highlighted in this paper indicates that this may be where AQC-style algorithms will most outperform classical alternatives.

The numerical analysis presented in Section \ref{sec:Results} offers an encouraging insight into how Hamiltonian simulation on higher dimensional instances may perform. The notion of mapping Euclidean distances into Hamiltonian energies is one that has many foreseeable applications in tackling lattice problems: there are similarities, for example, in the formulation of SVP$_\gamma$ and the approximate closest vector problem \cite{Peikert2016ACryptography}. Many lattice problems are closely related, meaning there are several areas one could apply the ideas laid out in this work.

Looking forward there are many interesting challenges to surmount. A major one is the issue of achieving better theoretical bounds on scaling complexity. One advantage of AQC is that time dependence relies on only one factor (minimum energy gap $\Delta$) meaning the source of time cost is easy to understand. In a sub-adiabatic regime, however, modelling eigenstate transitions probabilistically could be a natural progression for theoretical analysis of AQC-style algorithms. The development of quantum hardware that can realise this generalised Bose-Hubbard Hamiltonian and assume particle-particle offsite interactions is a target for experimental physicists, and generalising AQC-style algorithms to run on different hardware \textemdash such as coherent Ising machines \cite{Hamerly2016AConnections, Hamerly2018ExperimentalAnnealer} \textemdash will become increasingly investigated as progress continues towards a post-quantum world.

\begin{acknowledgments}
The authors would like to thank Adam Callison for helpful discussion. Alexandros Ghionis was supported through a studentship in the Quantum Systems Engineering Skills and Training  Hub  at  Imperial  College  London  funded  by  EP-SRC(EP/P510257/1).
\end{acknowledgments}

\appendix

\section{Qubit scaling}
\label{app:qubit scaling}
With these heuristic scaling assumptions we can derive the following analysis for the Single-Run algorithm (results for Multi-Run are similar). Using $m = cN$ for linear constant $c$, there are $m(N+1)$ particles in the system. Therefore the total particle number for Single-Run (denote $K_S$) is $K_S = cN^2 + cN$.

The Hilbert space size with $P$ particles distributed among $Q$ sites is 
\begin{equation}
\label{hilbert space}
    D = \frac{(P+Q-1)!}{(P)!(Q-1)!}.
\end{equation}
The qubit scaling equivalent is obtained by simply taking the base-two logarithm of the above expression, with values for $P,Q$ substituted in:
\begin{equation}
    D = \frac{(K_S+N)!}{K_S!N!} = \frac{(c N^2 + c N+N)!}{(c N^2 + c N)!N!}.
\end{equation}
By Stirling's approximation this is very close to 
\begin{equation}
    \frac{1}{\sqrt{2 \pi N}} \Big( \frac{e}{N} \Big)^N \bigg( \frac{(c N^2 + c N+N)!}{(c N^2 + c N)!} \bigg),
\end{equation}
which can be written 
\begin{equation}
\frac{1}{\sqrt{2 \pi N}} \Big( \frac{e}{N} \Big)^N \prod_{i=1}^{N} (c N^2 + c N + i).
\end{equation}
Bounding the product term, this is much less than 
\begin{equation}
    \frac{1}{\sqrt{2 \pi N}} \Big( \frac{e}{N} \Big)^N (c N^2 + (c + 1) N)^N = \frac{1}{\sqrt{2 \pi N}} ( e(c N + (c + 1) )^N.
\end{equation}
Leaving a system size in qubit terms of $\log_2 D$ bounded above by $O(N \log N)$. There are no analytical time bounds; this is an active research area in the community. What we can do is provide some analysis for a Grover search algorithm over the same solution space, giving us some post-quantum context for the complexity to be expected. Given that the solution space scales as $N^N = 2^{N \log N}$, search using Grover's algorithm scales as $2^{\frac{1}{2}N \log N}$.

\section{\label{app:band} Basis Band-diagonalisation}
In order to guarantee that offsite interaction terms $\gamma_{ij}$ in the problem Hamiltonian are nonzero only for small $|i-j|$ the basis must be altered to take a banded structure, but using only operations that preserve the basis. To demonstrate, consider the following example: The row span of the matrix
\begin{center}
$ \begin{bmatrix}
\textbf{a}\\
\textbf{b}\\
\textbf{c}
\end{bmatrix}
=
\begin{bmatrix}
a_{11} & a_{12} & 0 \\
0 & b_{22} & b_{23} \\
0 & 0 & c_{33}
\end{bmatrix} $
\end{center}
returns vectors $\textbf{v} = n_1 \textbf{a} + n_2 \textbf{b} + n_3 \textbf{c}$ of norms 
\begin{equation}
| \textbf{v} | = (n_1a_{11})^2 + (n_1a_{12} + n_2b_{22})^2 + (n_2b_{23}+n_3c_{33})^2.
\end{equation}
Expanding this vector norm and grouping into onsite terms and offsite terms gives
\begin{equation}
\begin{split}
| \textbf{v} | &= n_1^2(a_{11}^2 + a_{12}^2) + n_2^2(b_{22}^2 + b_{23}^2) + n_3^2(c_{33}^2) \\  &+ 2n_1n_2a_{12}b_{22} + 2n_2n_3b_{23}c_{33},
\end{split}
\end{equation}
and one can see that while there exist nearest-neighbour interaction terms ($\gamma_{12},\gamma_{23}$) there is no $\gamma_{13}$ term and so one can see how the banding structure is necessary to eliminate far away offsite particle-particle interactions.

Our solution is to iteratively eliminate elements far away from the leading diagonal using an argument that relies on taking the greatest common divisor ($gcd$) of as many elements as is needed to help us eliminate elements using Bezout's lemma.

Take a simple example to give a taste of what this algorithm is tapping into. Consider a prime determinant integer lattice, here taking the determinant to be some large $p$. Let us pick $p$ so that the lattice is in fact the row span of the following matrix. Every prime determinant integer lattice can be represented by a basis very similar in form to the one shown \cite{Goldstein2003OnPoints} \textemdash the pivot $p$ is not necessarily in the bottom right, but this does not affect the reduction:
\begin{equation}
 \begin{bmatrix}
\textbf{b}_1\\
\textbf{b}_2\\
\vdots \\
\textbf{b}_N
\end{bmatrix}
=
\begin{bmatrix}
1 & & & x_1 \\
 & \ddots & & \vdots \\
 & & 1 & x_{N-1} \\
  & & & p
\end{bmatrix} .
\end{equation}

For this simple case assume that the $gcd$ of $x_{i+1},x_{i+2}$ divides $x_i$. Then by Bezout's lemma there exists some $u,v,\delta$ such that 
\begin{equation}
\delta (u x_{i+1} + v x_{i+2}) = x_i.
\end{equation}
Now perform the lattice preserving row operations
\begin{equation}
\textbf{b}_i \rightarrow \textbf{b}_i^* = \textbf{b}_{i} - \delta u \textbf{b}_{i+1} - \delta v \textbf{b}_{i+2}.
\end{equation}
This should be performed iteratively from $i=1$ to $N-2$. After the first such iteration the above matrix looks as follows:
\begin{equation}
 \begin{bmatrix}
\textbf{b}_1^*\\
\textbf{b}_2\\
\textbf{b}_3 \\
\vdots \\
\textbf{b}_n
\end{bmatrix}
=
\begin{bmatrix}
1 & -\delta u & - \delta v & & & 0 \\
 & 1 & & & & x_2 \\
 & & 1 & & & x_3 \\
 & & & \ddots & & \vdots \\
 & & & & 1 & x_{n-1} \\
 & & & & & p
\end{bmatrix}.
\end{equation}

If $gcd(x_{i+1}, x_{i+2})$ does not divide $x_i$, then extend consideration to $x_{i+3}$, and so on until finding a group of numbers with $gcd$ dividing $x_i$. Each extra number means adding an extra band to the matrix so it is ideal to find a set of coprime (or with $gcd$ dividing $x_i$) entries $x_{i+1},...x_{i+j}$ for some small $j$. Fortunately, for $k$ randomly selected numbers the probability of them being coprime (a stronger condition than is needed) is $1/\zeta(k)$ which fast approaches one as $k$ increases. This means that even for high dimensional lattices one can be confident of tight bandings.

There are circumstances where $x_i$ is odd, and several of the entries below it are even. These are the only problematic cases where sticking strictly to the algorithm yields poor results. In these cases it is optimal to instead to multiply $\textbf{b}_i$ by $2$ so that $x_i$ can be eliminated efficiently. This is not ideal as it does not preserve the lattice, but instead increases the volume by a factor of $2$ (or $p$ if this is extended to some other small primes). What results is a basis for a sublattice. Fortunately these cases are rare enough that the mean volume increase on performing this algorithm over many lattices is very small.

To generalise this to any given HNF basis this procedure simply needs to be repeated for all dense columns, which will appear only above pivots that are not equal to $1$. While technically HNF bases can be dense in the upper triangle, this is not typical. Furthermore, HNF bases are particularly relevant cryptographically \cite{Micciancio2001ImprovingForm} as they afford a way of representing a bad basis that can be communicated with $O(N)$ key size (if dense this would be $O(N^2)$). Thus the property that makes this basis a good candidate for band-diagonalisation also makes it a good choice in terms of cryptographic efficiency.
\begin{figure}[H]
    \centering
    \includegraphics[width=\linewidth]{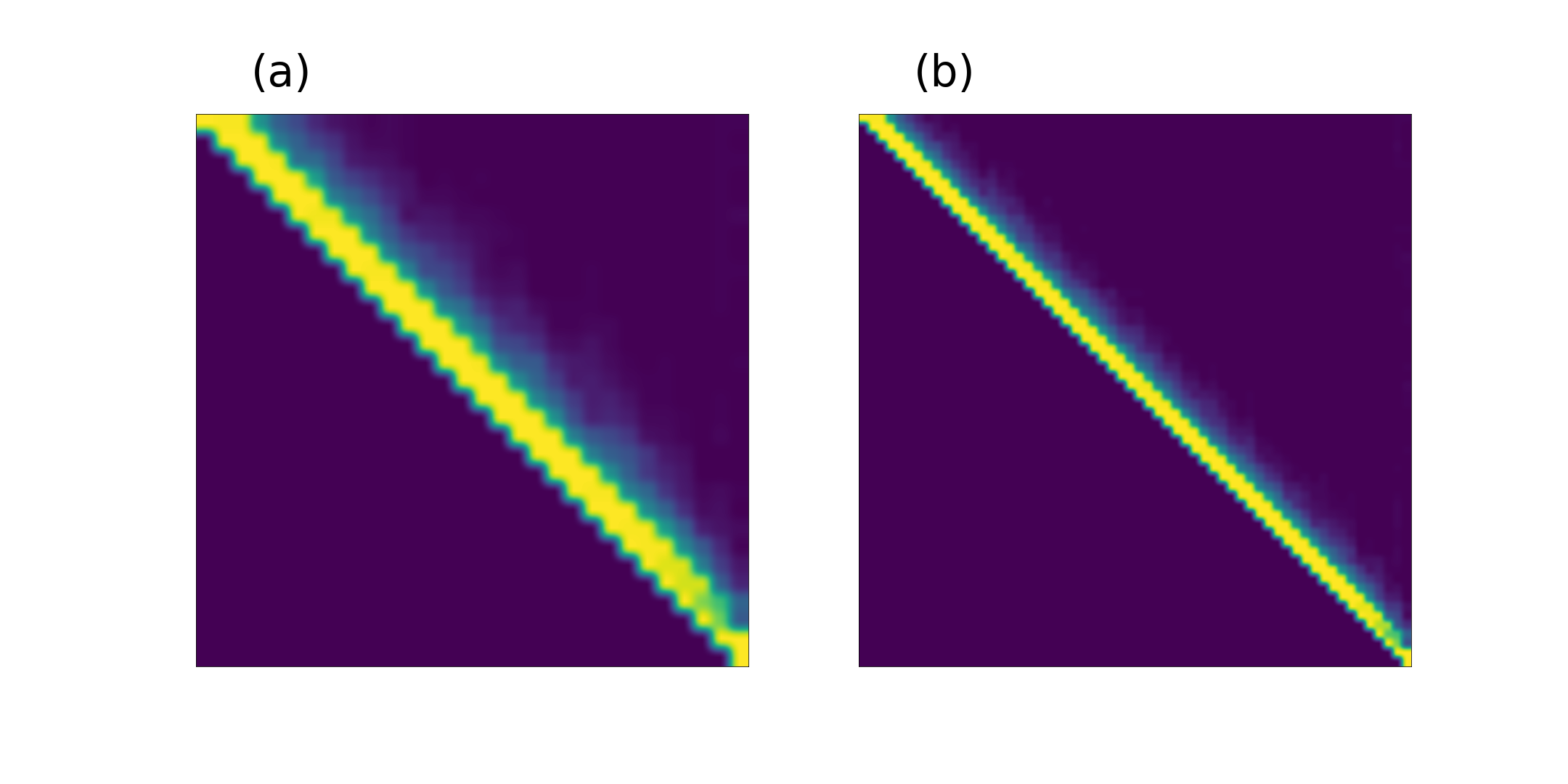}
    \caption{Banding algorithm performance over 100 instances of 30 dimensional lattices (a) and 60 dimensional lattices (b). The colour of each coordinate represents the average relative entry size.}
    \label{fig:tight_banding}
\end{figure}
Fig \ref{fig:tight_banding} shows the average results for the tight banding algorithm on HNF bases for 100 lattices in thirty and sixty dimensions, with the heat indicating the relative size of coefficients. The size of these coefficients reduces farther away from the leading diagonal. The dark blue squares represent zeros and so it is clear to see both the upper-triangular form of the row bases and also how the magnitude of the coordinates fades quickly to zeros above the leading diagonal. Moreover, increasing the dimension does not adversely affect the ability of the algorithm to produce a tightly banded lattice basis. The mean volume increase in thirty dimensions was 2.98 and in 60 dimensions was 7.99, meaning that while the algorithm tends not to preserve the lattice exactly, the volume of the basis is increased by a small factor. This is acceptable for solving SVP$_\gamma$. The effect of this band-diagonalisation algorithm is that in realising the quantum SVP algorithms described in this paper it is not necessary to consider particle-particle interaction terms for particles at sites which are far away from each other.

\section{Example run}
\label{sec:trivial_example}
Consider a very simple example to aide intuition in following the algorithm from lattice basis to final result and look at all the steps in between. The system comprises two particles in two lattice sites, with no offset ($m=0$). The Hilbert space is three dimensional and the Fock states are
\begin{equation}
\label{eq:ex_fock}
    \ket{02}, \ket{11}, \ket{20}.
\end{equation}
Take a good basis $\textbf{B}_G$, and a slightly longer bad basis $\textbf{B}_B$. Derive from $\textbf{B}_B$ the Gram matrix $\textbf{G}_B$ which defines coefficients for the problem Hamiltonian $H_P$
\begin{equation}
\label{eq:ex_algo}
\textbf{B}_G = 
\begin{pmatrix}
1 & 0\\ 0 & 2
\end{pmatrix}
\quad \textbf{B}_B = 
\begin{pmatrix}
1 & 2\\ 0 & -2
\end{pmatrix}
\quad \textbf{G}_B = 
\begin{pmatrix}
5 & -4\\ -4 & 4
\end{pmatrix}.
\end{equation}
Calculating coefficiants for $H_0$ from Eq \eqref{eq:tunnelling_Hamiltonian} and $H_P$ from Eq \eqref{eq:H_P} gives initial and final Hamiltonians of

\begin{equation}
    H_0 =
    \begin{pmatrix}
    0 & - \sqrt{2}  & 0 \\ - \sqrt{2} & 0 & - \sqrt{2}\\ 0 & - \sqrt{2} & 0
    \end{pmatrix},
    \quad
    H_P =
    \begin{pmatrix}
    16  & 0 & 0 \\ 0  & -8  & 0 \\ 0 & 0 & 20
    \end{pmatrix},\\
\end{equation}

and the time-dependent Hamiltonian matrix, using linear time evolution as per Eq \eqref{eq:linear evolution} is therefore
\begin{equation}
\label{eq:H_matrix}
    H(t) =
    \begin{pmatrix}
    16 \frac{t}{T} & - \sqrt{2} \Big( 1-\frac{t}{T} \Big) & 0 \\ - \sqrt{2} \Big( 1-\frac{t}{T}\Big)  & -8 \frac{t}{T} & - \sqrt{2} \Big( 1-\frac{t}{T}\Big) \\ 0 & - \sqrt{2} \Big( 1-\frac{t}{T}\Big) & 20 \frac{t}{T}
    \end{pmatrix}.
\end{equation}

The Hamiltonian ground state for $t=0$ is
\begin{equation}
\label{eq:ex_gstate}
    \ket{\psi_0} = \frac{1}{2}(\ket{02}+\sqrt{2}\ket{11}+\ket{20}).
\end{equation}
Initialise the system in this state and let it evolve. The probabilities of measuring the system to be in each of the Fock states during the evolution is as follows:
\begin{figure}[H]
    \centering
    \includegraphics[width=\linewidth]{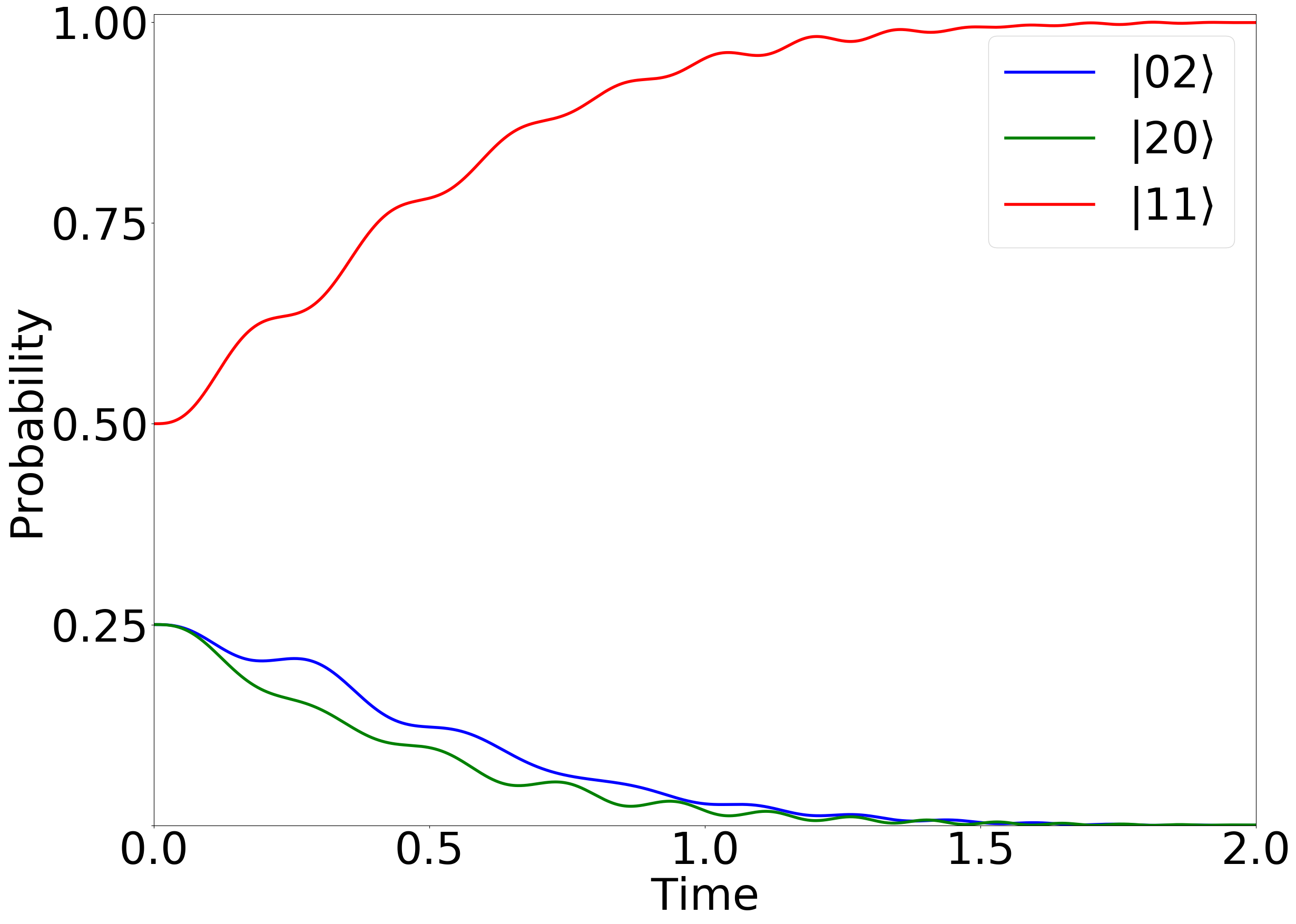}
    \caption{Each line represents the probability of finding the system in the corresponding Fock state as outlined in Eq \eqref{eq:ex_fock} as time progresses in the evolution.}
    \label{fig:mini_example}
\end{figure}
The Hamiltonian ground state for $t=T=2$ is
\begin{equation}
    \ket{\psi_T} = \ket{11},
\end{equation}
which is easy to see because
\begin{equation}
    \begin{pmatrix}
    1 & 1
    \end{pmatrix}
    \begin{pmatrix}
    1 & 2 \\
    0 & -2
    \end{pmatrix}
    = \begin{pmatrix}
    1 & 0
    \end{pmatrix},
\end{equation}
and $(1,0)$ is the shortest vector in the lattice. This can be seen by looking at $\textbf{B}_G$. This example would be one of many runs, each with a different number of particles, that together would constitute a Multi-Run algorithm to solve SVP for the lattice described by $\textbf{B}_B$ (and also $\textbf{B}_G$, though one would not have any prior knowledge of $\textbf{B}_G$).

\bibliography{references}

\end{document}